\documentclass[twocolumn, showpacs, preprintnumbers, amsmath, amssymb, superscriptaddress, pra]{revtex4}
\usepackage{graphicx}
\usepackage{dcolumn}
\usepackage{bm}
\usepackage{amssymb}

\begin{document}

\title{Tuning micropillar cavity birefringence by laser induced surface defects}

\author{Cristian Bonato}
\affiliation{Huygens Laboratory, Leiden University, P.O. Box 9504, 2300 RA Leiden, the Netherlands}

\author{Dapeng Ding}
\affiliation{Huygens Laboratory, Leiden University, P.O. Box 9504, 2300 RA Leiden, the Netherlands}

\author{Jan Gudat}
\affiliation{Huygens Laboratory, Leiden University, P.O. Box 9504, 2300 RA Leiden, the Netherlands}

\author{Susanna Thon}
\affiliation{University of California Santa Barbara, Santa Barbara, California 93106, USA}

\author{Hyochul Kim}
\affiliation{University of California Santa Barbara, Santa Barbara, California 93106, USA}

\author{Pierre M. Petroff}
\affiliation{University of California Santa Barbara, Santa Barbara, California 93106, USA}

\author{Martin P. van Exter}
\affiliation{Huygens Laboratory, Leiden University, P.O. Box 9504, 2300 RA Leiden, the Netherlands}

\author{Dirk Bouwmeester}
\affiliation{Huygens Laboratory, Leiden University, P.O. Box 9504, 2300 RA Leiden, the Netherlands}
\affiliation{University of California Santa Barbara, Santa Barbara, California 93106, USA}

\begin{abstract}
We demonstrate a technique to tune the optical properties of micropillar cavities by creating small defects on the sample surface near the cavity region with an intense focused laser beam. Such defects modify strain in the structure, changing the birefringence in a controllable way. We apply the technique to make the fundamental cavity mode polarization-degenerate and to fine tune the overall mode frequencies, as needed for applications in quantum information science.
\end{abstract}

\maketitle
Much work has been recently devoted to the development of semiconductor optical microcavities \cite {vahalaNature03} for quantum information processing applications. For example, a carefully designed cavity can be used to tailor the properties of single photon sources and to maximize their yield \cite{lounisRPP05, scheelJMO09, straufNP07}. In addition, quantum dots coupled to semiconductor microcavities provide a very promising system for the implementation of cavity quantum electrodynamics experiments \cite{reithmayerNature04, rakherPRL09}, and for hybrid quantum information protocols in which photons are used for long-distance transmission and matter qubits for local storage and processing \cite{ciracPRL97, vanenkPRL97}. However, some technical issues are yet to be solved: among these, the fine tuning of the microcavity optical properties. To generate quantum superpositions and to exploit quantum interference effects, which are at the heart of quantum information protocols, the states which form the superposition must be indistinguishable. In other words, if the polarization degree of freedom encodes the quantum bit, there must be no way to obtain information about its polarization by observing other degrees of freedom. Therefore, the implementation of a quantum interface between the polarization state of a single photon and a two-level system requires the cavity mode to be polarization-degenerate.  A second problem is that the cavity resonance frequency and the frequency of the two-level system, in our case semiconductor self-assembled quantum dots, must be matched with a precision which is currently impossible to obtain deterministically in the fabrication process. Several frequency tuning techniques are commonly used (like temperature tuning \cite{reithmayerNature04} or Stark shift \cite{hogelePRL04}) but they are temporary and leave the cavity birefringence unchanged.

Here we demonstrate an all-optical technique, originally developed to tailor the polarization properties of vertical-cavity surface-emitting lasers \cite{doornAPL96a, doornAPL96b}, to apply a controlled and permanent birefringence to the optical micropillar cavities. In this way the frequency shift of the two polarization modes can be tuned at will, allowing polarization-degeneracy. We will show that this technique permits control of the frequencies of the two polarization modes almost independently of one another, providing a tuning range of a few Angstroms.

The technique is based on the creation of a permanent defect on the surface of the sample near the cavity by means of a strongly focused laser beam (see Fig.~\ref{Figure }, upper figure). The sample locally melts, creating a hole with some material accumulated on the edges. Such holes are $3-5 \mu m$ wide, with depth varying from 30 nm to 2 $\mu m$ depending on the burning time, and affect the strain (and therefore the birefringence) in the structure. The magnitude of the induced stress can be varied by tuning the laser power and the exposure time, while its orientation is determined by the position of the burn around the cavity. In our experiments, the defect is created by a Ti-sapphire laser (about 250 mW power) tuned to 770 nm in order to have sufficient absorption by the semiconductor material, tightly focused on the structure by a high numerical aperture (NA) aspheric lens $L_1$ (focal length $f_0 = 4.02$ mm, $NA = 0.6$). The burn is precisely positioned onto the sample by means of an optical system consisting of the focusing lens $L_1$ and a second lens $L_2$ (focal length $f = 150$ mm) which images the sample onto a charge-coupled device (CCD) camera (placed in the focal plane of the lens $L_2$).

\begin{figure}[p]
\caption{\label{Figure 1} \emph{(a)} Hole burning locations for polarization-splitting compensation \emph{(b)} Atomic force microscpe (AFM) image of a hole. \emph{(c)} Compensation of the polarization splitting for the fundamental cavity mode.}
\includegraphics[width=8 cm] {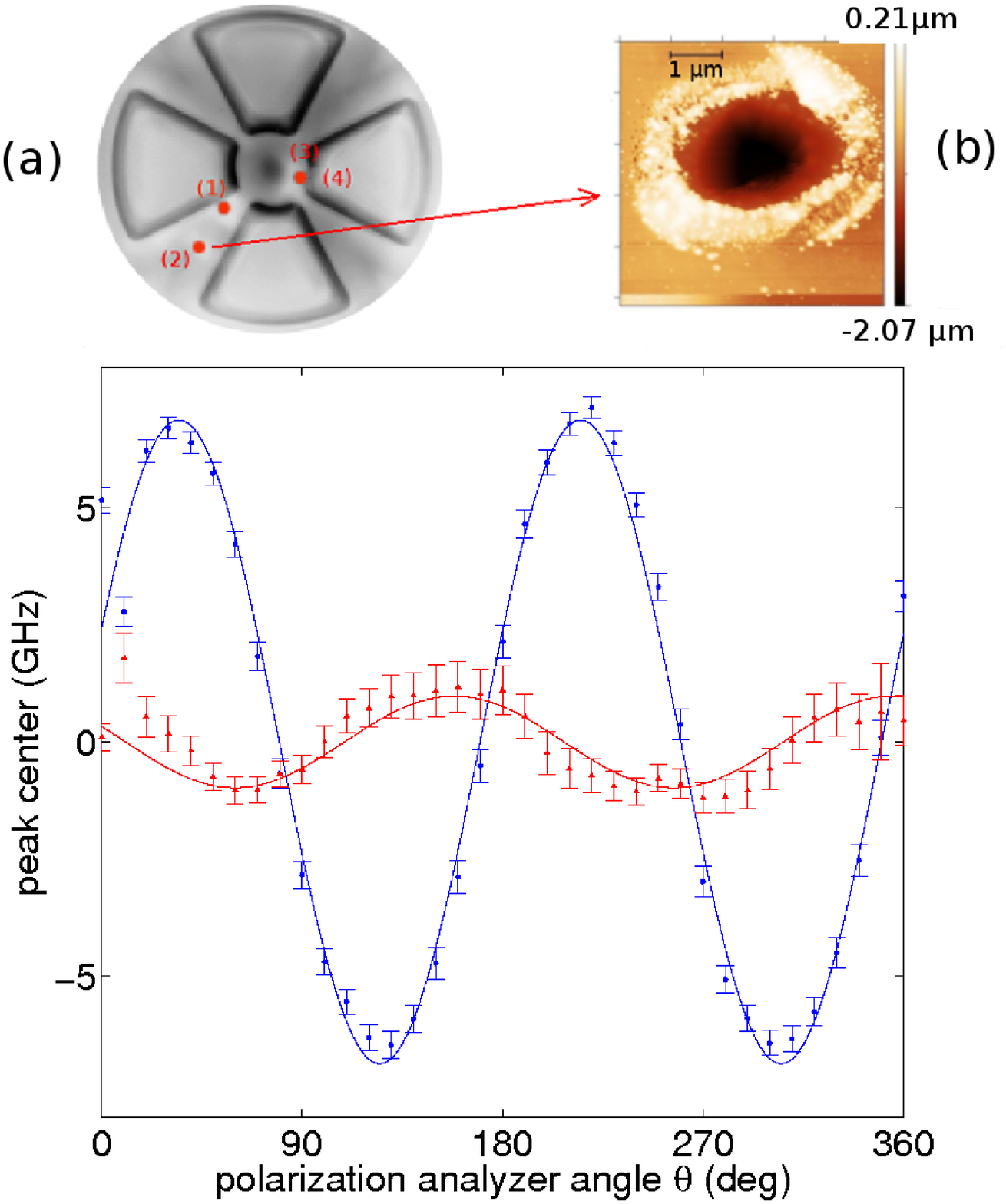}
\end{figure}

The micropillar samples we investigated were grown by molecular-beam epitaxy on a GaAs [100] substrate. The microcavity consists of two distributed Bragg reflector (DBR) mirrors made by alternating layers of GaAs and Al$_{0.9}$Ga$_{0.1}$As (one-quarter optical thickness, $32$ pairs for the bottom DBR, $4.8 \mu m$ thick, and $23$ pairs for the top DBR, $3.5 \mu m$ thick), spaced by a $\lambda$-thick GaAs cavity layer with embedded InGaAs/GaAs self-assembled quantum dots. Trenches are etched through the sample ($4.3 \mu m$ thick, down through the active region), and the sample is placed in an oxidation furnace to create an oxidation aperture in the AlAs layer which provides gentle lateral confinement of the optical mode \cite{stoltzAPL05}. The micropillar structures, typically $30 \mu m$ in diameter, are very robust (see \cite{straufNP07} for a 3D sketch). The oxidation aperture determines a mode waist of about $1-2 \mu m$ at the center of the structure. In this paper, we will focus on the properties of the fundamental transverse cavity mode, which exhibits a very good spatial Gaussian shape and is split into two orthogonally-polarized submodes ($M_A^{[00]}$ and $M_B^{[00]}$). The defects, burnt a few microns away from the center of the micropillar, do not reduce the optical quality of the cavity.

\begin{figure}[p]
\caption{\label{Figure 2} Optical microscope image of a micropillar cavity with holes (labeled 1-24) burnt chronologically on the structure in the regions between the trenches.}
\includegraphics[width=6 cm] {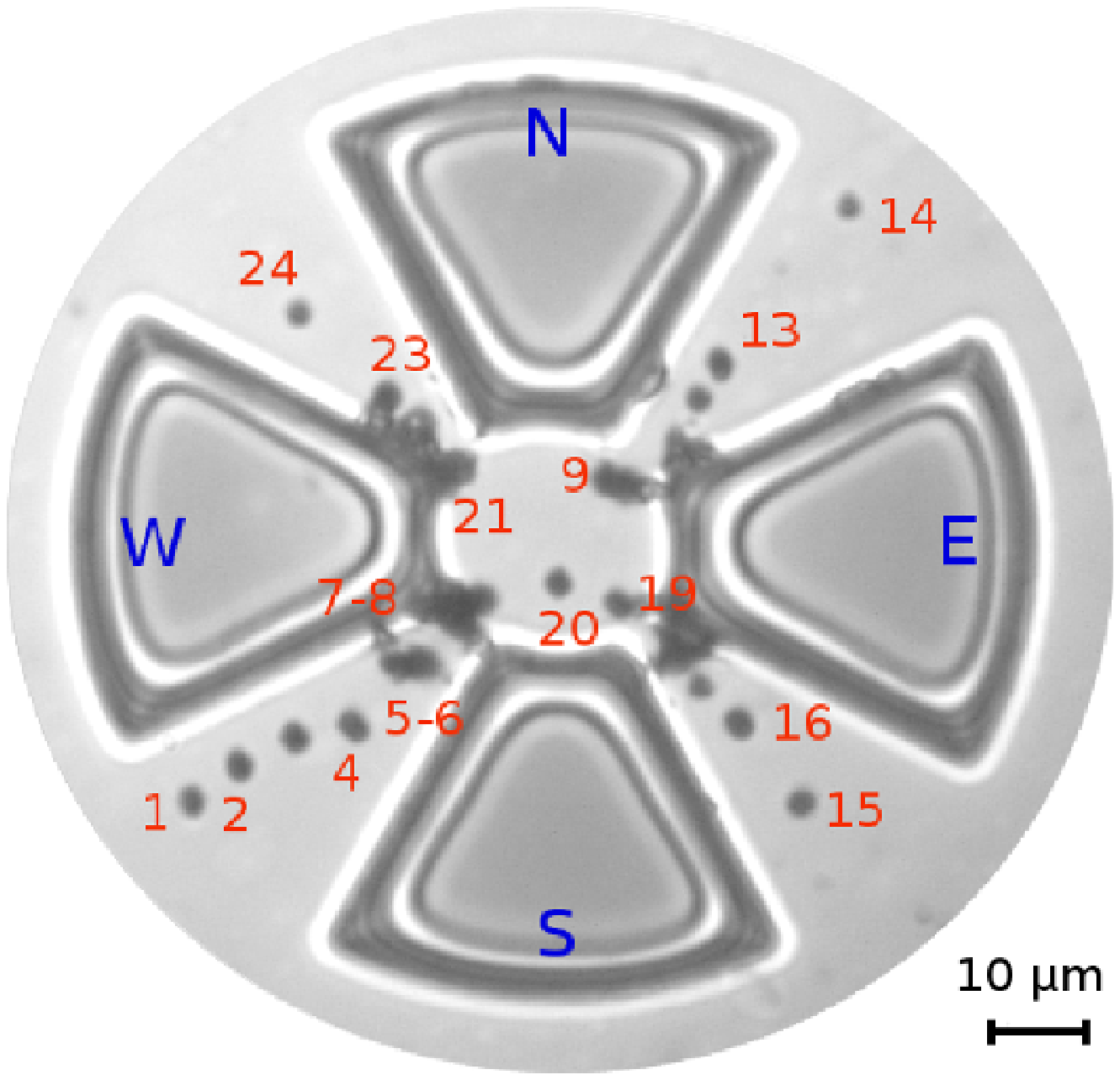}
\end{figure}

The spectrum of the cavity modes can be characterized by pumping the semiconductor material above the band-gap with a Ti-Sapphire beam of a few $mW$ and observing the cavity-shaped photoluminescence on a spectrometer (resolution $5.5$ GHz/pixel) equipped with a CCD array. Its polarization dependence is characterized by placing an analyzer consisting of a fixed linear polarizer and a rotating half-wave plate in front of the spectrometer, so that the polarization state in the spectrometer is constant and the measurements are not affected by the polarization response of the grating.

The spectral splitting of the two polarization modes is typically around few GHz, which is smaller than the spectral width of the mode and comparable to the spectrometer resolution. A direct measurement of the centers of the two overlapping modes is therefore not possible. However, the two modes are orthogonally polarized, so that $M_A^{[00]}$ ($M_B^{[00]}$), centered at frequency $\nu_A$ ($\nu_B$) dominates completely at some analyzer angle $\theta_A$ ($\theta_A+\pi/2$). Rotating the analyzer, the center of the peak shifts periodically between $\nu_A$ and $\nu_B$. The accuracy in the determination of the position of the peak can be greatly enhanced beyond the spectrometer resolution by means of a Lorentzian fit of the peak. The periodic oscillations of the peak center as a function of the analyzer angle can be clearly resolved, as shown in Fig. 1. Experimental data for a cavity with no holes burnt are shown by the blue curve: the separation between the central frequencies of the two polarization modes is measured to be $\Delta \nu = 13.7 \pm 0.3$ GHz. The FWHM of the fundamental mode peak, measured selecting one single polarization mode, is $30.1 \pm 0.4$ GHz.

\begin{figure}[p]
\caption{\label{Figure 3} Resonance frequency change when holes are burnt on the cavity shown in Fig. 2. Insets show two extreme spectra.}
\includegraphics[width=15 cm] {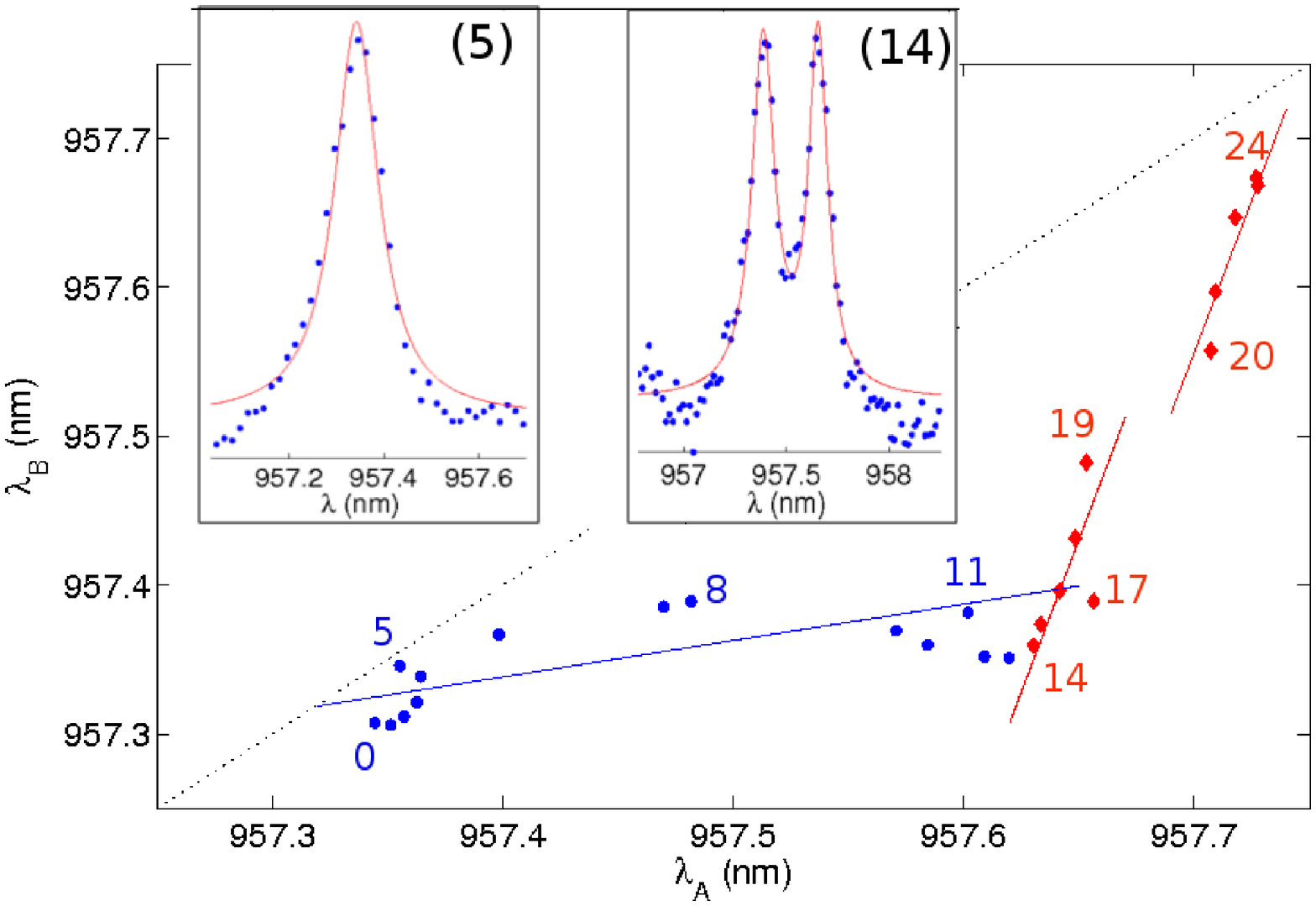}
\end{figure}

Next we modified the cavity by burning holes at locations chosen on the basis of the expected angular and $1/r$ dependence of the induced strain \cite{doornIEEE98} ($r$ being the distance of the burn from the cavity center) and of initial tests performed on different cavities. After burning one hole for one minute on the surface between the southern and western trenches (hole (1) on the inset in Fig. 1), close to the cavity center, the splitting is reduced from $13.7 \pm 0.3$ GHz to $6.1 \pm 0.7$ GHz. A second hole (2) did not lead to a strong reduction. After burning two more holes on the edge of the eastern trench (holes (3) and (4), each one for one minute), the splitting is reduced to $\Delta \nu = 2.0 \pm 0.2$ GHz.

More tests were performed on a new cavity in order to understand how the spectrum of the two peaks changes when a hole is burnt at different positions (see Fig.~2). The results are illustrated in Fig. 3, where the central wavelengths of the two polarization peaks ($\lambda_A$ and $\lambda_B$) are plotted on the horizontal and vertical axes. Each point on the graph corresponds to a hole burnt on the sample. First, eight holes were burnt between the southern and western trenches, starting farther from and moving closer to the cavity centers, and then six more holes were burnt between the northern and the eastern trenches continuing in the same direction (along the $[001]$ crystal lattice orientation). The effect of the first few holes is to reduce the spectral splitting of the two polarization modes, reaching a minimum of $\Delta \nu = 3.2 \pm 0.4$ GHz. After that, $\lambda_B$ remains approximately constant, while $\lambda_A$ increases. The closer the holes are burnt with respect to the center, the larger the frequency shift. Due to the combined effect of the $14$ holes burnt along this direction, $\Delta \nu$ becomes $108 \pm 1$ GHz and the two peaks can be clearly resolved (inset in Fig.~3). Burning holes along the orthogonal direction, $\lambda_B$ changes much faster than $\lambda_A$. By adjusting the hole orientation and distance from the center, we have a way to tune almost independently the frequency of the two polarization modes with a range of about $100-150$ GHz.

This effect can be explained by a simple model, based on the tensorial relationship between stress, strain and the optical properties of the material (see Appendix A). Essentially, the anisotropic component of the stress changes the splitting and the isotropic component affects the absolute frequencies of the two submodes. In particular, we find that for $N$ holes burnt along the $x$ direction:
\begin {equation}
n_B^{(N)} = n_0  + \left( n_A^{(N)}-n_0 \right) \left( -\frac{\Pi_2}{\Pi_1} \right)
\end {equation}
which corresponds to a straight line with slope $\rho_1 = -\Pi_2/\Pi_1$. $\Pi_1$ and $\Pi_2$ are quantities which depend on the tensorial elastic ($C_{ij}$) and elasto-optic ($p_{ij}$) coefficients of the material ($\Pi_1 = p_{11} C_{11}-p_{12}C_{12}$ and $\Pi_2 = p_{11} C_{12}-p_{12}C_{11}$). For the holes burnt along the $y$ direction we find a similar linear relationship with inverted slope $\rho_2 = -\Pi_1/\Pi_2$. As can be seen in Fig.~3, the data fit quite well with this model: the fitted slope for the lines is $4.0 \pm 0.5$. Literature values for the bulk GaAs and AlAs thermal and elasto-optic properties \cite{watsonJLT04} give a slope of around $1.5$. However, we do not expect these values to be perfectly compatible since our model does not take into account the bimorphic structure formed by the oxidized AlAs layer and by the DBR mirrors.

The cavity resonance frequency and polarization splitting for holes burnt at room temperature were measured at $4$ K, showing values significantly different from the room-temperature ones. We repeated the hole-burning process directly at low-temperature, increasing the burning laser power to $500$ mW ($532$ nm). Polarization degeneracy could be achieved as well as frequency tuning, albeit with a different slope $\rho_2 = 1.2 \pm 0.5$. The stability of the effects was tested by warming up and cooling down the device a few times: a difference of the order of $10\%$ was found for the first cooldown after burning (consistent with the results in \cite{doornAPL96b}), while the deviation in the splitting is within $1-2$ GHz for the successive cool-downs.

In conclusion, we introduced a technique to permanently tune the polarization and spectral properties of optical micropillar cavities. By laser-burning a small defect on the sample surface near the cavity, we can induce a controllable amount of birefringence in the structure. By adjusting the position of the defect, we control the central wavelengths of the two polarization submodes of the fundamental cavity mode. This technique enables the implementation of polarization-degenerate semiconductor micropillars for quantum information processing and it may find applications for fine tuning of other kinds of semiconductor microcavities whose optical properties are influenced by material strain, such as photonic crystal defect cavities and microdisk cavities.

\begin{acknowledgments}
This work was supported by the NSF grant 0901886, and the Marie-Curie No. EXT-CT-2006-042580. We thank Brian Ashcroft for the AFM images and Andor for the CCD camera.
\end{acknowledgments}

\appendix
\section {Model for cavity resonance tuning via hot-spot burning}
Here we describe a model to explain the effect of hot-spot burning on the modes of micropillar cavities. In particular, we extend the model developed in \cite{doornAPL96a, doornAPL96b} to account for isotropic strain.\\
Consider a cavity with a symmetry axis along $(0,0, z)$ and suppose that the $x$ and $y$ axes are oriented along the crystal lattice axes of a cubic crystal. Stress in the structure is described by the stress tensor:
\begin{equation}
\sigma = \left[
\begin{array}{lll}
\sigma_{xx} & \sigma_{xy} & \sigma_{xz}\\
\sigma_{yx} & \sigma_{yy} & \sigma_{yz}\\
\sigma_{zx} & \sigma_{zy} & \sigma_{zz}
\end{array}
\right]
\end{equation}

Suppose now that a hole is burnt at the position $P (x_0, y_0, 0)$. Thermal expansion caused by point source heating induces stress, which can be calculated as (see \cite{doornAPL96b}):
\begin {equation}
\sigma_{ij} = \gamma \frac{A}{r} \left[  \delta_{ij}-\frac{(x_i-x_i^{(0)}) (x_j-x_j^{(0)})}{r^2}\right] \qquad i,j = 1,2,3
\end{equation}
where $\gamma = \alpha (C_{11}-C_{12})(C_{11}+2C_{12})/6C_{11}$, $C_{ij}$ are the crystal elastic constants and $\alpha$ is the thermal expansion coefficient. In our experiments, we induce a plastic permanent deformation of the structure, which we assume to have the same form as the stress caused by heating.\\
The applied stress induces strain in the crystal lattice, as described by:
\begin{equation}
 \sigma_{ij} = C_{ijkl} u_{kl}
\end{equation}
where $\lbrace c_{ijkl} \rbrace$ and $\lbrace u_{kl} \rbrace$ are respectively the components of the elastic stiffness tensor and of the strain tensor.
The lattice strain induces birefringence in the structure, described by a perturbation to the dielectric impermeability tensor $B$.
This is known as the photoelastic effect, described by the elasto-optic tensor ${p_{ijkl}}$:
\begin {equation}
B_{ij} = p_{ijkl} u_{kl}
\end{equation}

In order to simplify the model we can make the following approximations:
\begin{itemize}
  \item the cavity is small and holes are burnt quite far from it, so that in the cavity region the perturbation to the permittivity tensor induced by the stress is homogeneous.
  \item the cavity is thin in the $z$ direction, so that we can assume no $z$-dependence of the permittivity tensor. In this case, everything reduces to 2-by-2 tensors.
  \item we neglect the spatial inhomogeneity of the cavity, which gives the Hermite-Gaussian transverse modes, and we just concentrate on the longitudinal modes.
\end{itemize}

The permittivity tensor can be diagonalized to find the principal axes (characterized by refractive indices $n_1$ and $n_2$). We assume that the fundamental mode has two orthogonally-polarized submodes, oriented along the anisotropy principal axes, whose wavelength is proportional to the refractive index of the axis.

The perturbation to the dielectric impermeability tensor is:
\begin {equation}
\delta B = \left[
\begin{array}{cc}
  \delta B_{xx} &   \delta B_{xy} \\
  \delta B_{xy} &   \delta B_{yy}
\end{array}
\right]
\end{equation}
where, for a $4\bar{3}m$ cubic crystal, $\delta B_{xx} = c_0 [\Pi_1 \sigma_{xx} - \Pi_2 \sigma_{yy}]$, $\delta B_{yy} = c_0 [- \Pi_2 \sigma_{xx} + \Pi_1 \sigma_{yy}]$, $\delta B_{xy} = \frac{p_{44}}{C_{44}} \sigma_{xy}$, $c_0^{-1} = (C_{11}-C_{12})(C_{11}+2C_{12})$, $\Pi_1 = p_{11} C_{11}-p_{12}C_{12}$ and $\Pi_2 = p_{11} C_{12}-p_{12}C_{11}$. Since $B_i = 1/n^2_i$, in the case of a small perturbation:
\begin {equation}
\Delta n_i \sim -\frac{1}{2} n_0^3 \Delta B_i
\end {equation}
For a hole burnt a distance $r$ from the cavity center and oriented at an angle $\theta$, Eq. (2) gives:
\begin {equation}
\begin{split}
\sigma_{xx} &= \gamma \frac{A}{r} \sin^2 \theta \\
\sigma_{yy} &= \gamma \frac{A}{r} \cos^2 \theta \\
\sigma_{xy} &= -\gamma \frac{A}{r} \sin \theta \cos \theta\\
\end{split}
\end {equation}

\subsection {Cavity birefringence}
In order to tune the birefringence of the cavity, only the anisotropic component of the tensor $B$ is important. As described in \cite{doornAPL96b}, the birefringence angle is:
\begin {equation}
\tan 2\phi = \frac{2 \delta B_{xy}}{\delta B_{xx}-\delta B_{yy}} = \beta \tan 2\theta
\end{equation}
where:
\begin {equation}
\beta = \frac{2 p_{44}}{p_{11}-p_{12}} \frac{C_{11}-C_{12}}{2C_{44}} \frac{C_{11}+2 C_{12}}{C_{11}+C_{12}}
\end{equation}

\subsection {Cavity resonance tuning}
To tune the cavity resonance we are interested in the absolute wavelengths of the modes, and we need also the isotropic components of the involved tensors. Burning holes along one of the crystal axes ($\theta = 0$):
\begin {equation}
\delta B =  \frac{2\xi}{r}\left[
\begin{array}{cc}
  \Pi_1 &  0\\
  0 & -\Pi_2
\end{array}
\right]
\qquad
\xi = \frac{\alpha A}{12 C_{11}}
\end{equation}
which means $\Delta n_1 = -\xi n_0^3 \Pi_1/r$ and $\Delta n_2 = \xi n_0^3 \Pi_2/r$. For the $j$-th hole along the $\theta = 0$ direction, the change in the refractive index is:
\begin {equation}
(\delta n_1)_j = \xi n_0^3 \frac{\Pi_1}{r_j} \qquad (\delta n_2)_j = \xi n_0^3 \frac{\Pi_2}{r_j}
\end {equation}
The contribution of $N$ holes gives the following values:
\begin {equation}
n_1^{(N)} = n_0 + \Gamma_N n_0^3 \Pi_1 \qquad n_2^{(N)} = n_0 - \Gamma_N n_0^3 \Pi_2
\end{equation}
where:
\begin {equation}
\Gamma_N = \xi \left( \sum_{j=1}^N \frac{1}{r_j} \right)
\end{equation}
Eliminating $\Gamma_N$, a relationship between the refractive indices of the two modes after burning $N$ holes can be found:
\begin {equation}
n_2^{(N)} = n_0  + \left( n_1^{(N)}-n_0 \right) \left( -\frac{\Pi_2}{\Pi_1}\right)
\end {equation}
which corresponds to a straight line with slope $\rho_1 = -\frac{\Pi_2}{\Pi_1}$.

Burning holes along the orthogonal direction ($\theta = \pi/2$), $\Delta n_1 = -\xi \frac{n_0^3}{r} \Pi_2$ and $\Delta n_2 = \xi \frac{n_0^3}{r} \Pi_1$, we find:
\begin {equation}
n_2^{(N)} = n_0  + \left( n_1^{(N)}-n_0 \right) \left( -\frac{\Pi_1}{\Pi_2} \right)
\end {equation}
which corresponds to a line with slope $\rho_2 = -\frac{\Pi_1}{\Pi_2}$, so that $\rho_1 \cdot \rho_2 = 1$.\\

Literature values for bulk material parameters are not very helpful in our case, since we are dealing with a complicated structure and alloys. However, just to get a sense of the effect, parameters for bulk GaAs and AlAs \cite{watsonJLT04} are reported in the following Table (the values for the $C_{ij}$ coefficients are in units of $10^{10} Nm^{-2}$):\\
\\
\begin{center}
\begin{tabular}{|c|c|c|c|c|c|c|c|}
  \hline
       & $C_{11}$ & $C_{12}$ & $C_{44}$ & $p_{11}$ & $p_{12}$ & $p_{44}$ \\
  \hline
  GaAs & 11.88 & 5.38 & 5.94 & -0.165 & -0.140 & -0.072 \\
  AlAs & 12.02 & 5.70 & 5.89 & -0.040 & -0.035 & -0.010 \\
  \hline
\end{tabular}
\end{center}

\section {AFM images of some holes burnt on the sample}
In the following we attach some higher-resolution images of the laser-induced defects taken with the atomic-force microscope.

\begin{figure}[p]
\caption {AFM images of holes burnt on the sample}
\includegraphics[width=10 cm] {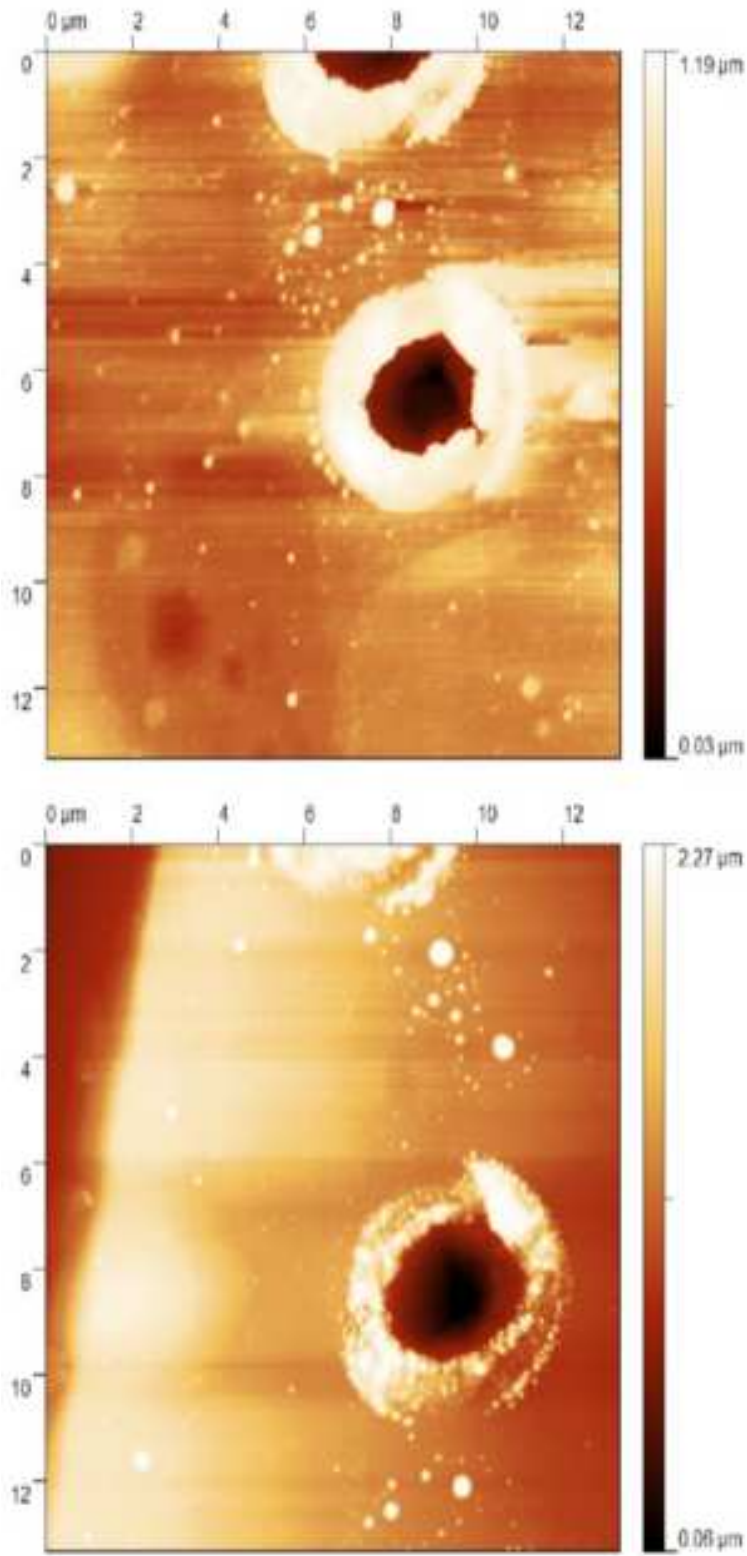}
\end{figure}

\end{document}